# Gate-tunable anomalous Hall effect in stacked van der Waals ferromagnetic insulator – topological insulator heterostructures

A. E. Llacsahuanga Allcca[1,2], X. C. Pan[3], I. Miotkowski[1], K. Tanigaki[3,4] and Y. P. Chen[1,2,3,5,6,7]

[1] *Department of Physics and Astronomy, Purdue University, West Lafayette, Indiana 47907, USA*

[2] *Purdue Quantum Science and Engineering Institute and Birck Nanotechnology Center, Purdue University, West Lafayette, Indiana 47907, USA*

[3] *WPI Advanced Institute for Materials Research (AIMR), Tohoku University Sendai 980-8577, Japan*

[4] *Beijing Academy of Quantum Information Sciences, Beijing 100193, China*

[5] *School of Electrical and Computer Engineering, Purdue University, West Lafayette, Indiana 47907, USA*

[6] *Institute of Physics and Astronomy and Villum Center for Hybrid Quantum Materials and Devices, Aarhus University, 8000 Aarhus-C, Denmark*

[7] *Center for Science and Innovation in Spintronics, Tohoku University Sendai 980-8577, Japan*

[*]aallcca@purdue.edu; yongchen@purdue.edu

**Abstract**

The search of novel topological phases, such as the quantum anomalous Hall insulator (QAHI) or the axion insulator, has motivated different schemes to introduce magnetism into topological insulators. One scheme is to introduce ferromagnetic dopants in topological insulators. However, it is generally challenging and requires carefully engineered growth/heterostructures or relatively low temperatures to observe the QAHI due to issues such as the added disorder with ferromagnetic dopants. Another promising scheme is using the magnetic proximity effect with a magnetic insulator to magnetize the topological insulator. Most of these heterostructures are synthesized so far by growth techniques such as molecular beam epitaxy and metallic organic chemical vapor deposition. These are not readily applicable to allow mixing and matching many of the available ferromagnetic and topological insulators due to difference in growth conditions and lattice mismatch. Here, we demonstrate that the magnetic proximity effect can still be obtained in stacked heterostructures assembled via the dry transfer of exfoliated micrometer-sized thin flakes of van der Waals topological insulator and magnetic insulator materials ($BiSbTeSe_2/Cr_2Ge_2Te_6$), as evidenced in the observation of an anomalous Hall effect (AHE). Furthermore, devices made from these heterostructures can allow modulation of the AHE when controlling the carrier density via electrostatic gating. These results show that simple mechanical transfer of magnetic van der Waals materials provides another possible avenue to magnetize topological insulators by magnetic proximity effect, a key step towards further realization of novel topological phases such as QAHI and axion insulators.

Keywords: Magnetic proximity effect, topological insulator, 2D ferromagnetic insulators, van der Waals heterostructures, anomalous Hall effect.

In the attempts for the realization of new magnetic topological states, extrinsic (doping [Chang2013], interfacing [Yao2019, Mogi2019, Alegria2014, Chong2018, Nagata2021]) and intrinsic (magnetic topological insulators [Liu2020] or correlated Chern insulators in twisted bilayer graphene [Sharpe2019, Chen2020, Polshyn2020, Stepanov2021]) methods of integrating magnetism have been employed. Among them it can be considered that engineering the interface holds the promise to be a more versatile platform for exploration, for example the magnetic material and topological insulator can be chosen from a wide inventory of materials to create the heterostructures. The main limitation on interface engineering depends on the method to create the interface, for example, growing the materials on top of each other requires being mindful of lattice mismatch and chemical diffusion. Previously, this has been explored using metal-organic chemical vapor deposition (e.g., $Cr_2Ge_2Te_6$/$Bi_2Te_3$ [Alegria2014]) and molecular beam epitaxy (MBE) (e.g. $EuS$/$Bi_2Se_3$ [Wei2013], $Y_3Fe_5O_{12}$/$Bi_2Se_3$ [Lang2014], $Y_3Fe_5O_{12}$/$(Bi_xSb_{1-x})_2Te_3$ [Jiang2015], $Tm_3Fe_5O_{12}$/$(Bi_xSb_{1-x})_2Te_3$ [Tang2017], $Eu_3Fe_5O_{12}$/$(Bi,Sb)_2Te_3$ [Zou2022], $Cr_2Ge_2Te_6$/$(Bi_xSb_{1-x})_2Te_3$ [Yao2019, Mogi2019, Mogi2021], $(Zn,Cr)Te$/$(Bi,Sb)_2Te_3$/$(Zn,Cr)Te$ [Watanabe2019]). More recently, wet transfer (e.g., $Y_3Fe_5O_{12}$/$Bi_2Se_3$ [Che2018], $Cr_2Ge_2Te_6$/$BiSbTeSe_2$ [Nagata2021]) and dry transfer ($Cr_2Ge_2Te_6$/$BiSbTeSe_2$ [Chong2018]) have also been used to assemble the heterostructures using layered or van der Waals (vdW) materials that can help relaxing the requirements such as lattice matching for interface engineering. In principle, the weak interlayer coupling in vdW materials makes possible the use of simple mechanical transfer to create an interface free from the issues or limits in growth techniques. In this work, we prepare van der Waals heterostructures using such mechanical dry transfer method. By choosing an appropriate bulk-insulating topological insulator along with an insulating van der Waals ferromagnet we show that the heterostructures of these materials exhibits an anomalous

Hall effect (AHE). Importantly and in contrast to previous similarly-fabricated (wet and dry transfer) samples, our observed AHE shows clear hysteresis. Furthermore, due to the bulk-insulating nature of the chosen topological insulator it is possible to change the carrier density of the surface states in the heterostructure by electrostatic gating. This is shown to be able to modulate the amplitude of the AHE (while its sign remains unchanged) following a similar trend as that of the longitudinal resistance of the device when tuning the back gate voltage. These results are interpreted as a magnetic proximity effect induced in the top surface states of the topological insulator due to the ferromagnetic insulator, realized in a fully vdW heterostructure bypassing growth constraint and limitations found in previous works.

We use thin flakes mechanically exfoliated from the vdW layered $Cr_2Ge_2Te_6$ (CGT) as the ferromagnetic insulator in our heterostructure. Single crystals of this material were grown via a self-flux technique as described in a previous work [Idzuchi2019]. For the topological insulator we employ flakes exfoliated from a single crystal of $BiSbTeSe_2$ (BSTS) grown by the vertical Bridgman technique as described in ref. [Xu2014]. Taking advantage of the layered nature of these crystals, it is possible to assemble heterostructures while avoiding those defects and irregularities in the interface that might occur in heterostructures assembled via growth due to lattice mismatch and chemical diffusion (both issues can be detrimental for AHE observation) [Yao2019, Bhattacharyya2021]. We chose topological insulator flakes with thickness in the range of 40 – 70 nm, which maintain the intrinsic insulating bulk and conduct mainly through the surface states [Xu2014, Xu2019, Chong2019]. For the ferromagnetic insulator, we chose flakes with thickness in range of 4 – 10 nm, whose low temperature magnetization shows a more rectangular hysteresis loop with a clear coercivity and higher remnant magnetization compared to the smooth magnetization behavior for bulk CGT and the more complicated hysteresis loops with softer magnetic behavior in thicker CGT flakes (>10 nm). The latter are usually attributed to formation of labyrinth type domains [Jagla2005, Fei2018] (see figure S1 for an evolution of hysteresis loops of CGT as thickness is increased, measured using Magneto Optics Kerr Effect). This distinct magnetic hysteresis loop behavior was expected to be inherited into the topological insulator and would make the AHE more distinguishable from other nonlinear Hall effects. In figure 1a and 1b, a schematic of the heterostructures and the microscope image of a typical device are shown, respectively. As it can be seen, the CGT flake did not fully cover the topological insulator. This was done to ease the contact fabrication, as CGT is highly insulating at low temperatures (see figure S2) and a full coverage of the BSTS flakes would lead to poor contacts. In figure 1c, ambipolar field effect in the longitudinal resistance measured by the four-probe method can be observed, showcasing that the heterostructure still allows similar gate tuning of the resistance as in BSTS-only devices [Xu2014]. An out-of-plane magnetic field is applied and swept to extract the Hall resistance in the samples. Figure 1d shows representative traces of the Hall resistance and longitudinal resistance vs. magnetic field for the device. As seen in the figure, a clear AHE with a rectangular hysteresis loop with an amplitude of few ohms and coercive field of ~0.035 T is observed in the $R_{xy}$ curve, indicating that a magnetization in the conducting carriers is introduced by interfacing the topological insulator BSTS with the magnetic insulator CGT. Furthermore, the longitudinal resistance ($R_{xx}$) of the device shows a minimum in the magnetoresistance around zero field, typically attributed to the weak antilocalization (WAL) behavior due to the strong spin orbit coupling in the topological insulator. Phenomenologically, the magnetoresistance can be fitted to the Hikami-Larkin-Nagaoka equation used to describe the WAL and how the fitting parameters varies with the back gate (see figure S3) is similar to the WAL behavior reported in the literature [Steinberg2011, Jauregui2015].

To further study the hysteresis loop (attributed to AHE) in $R_{xy}$, the main focus of this paper, we measure $R_{xy}$ as the function of the gate voltage ($V_{BG}$, applied to the silicon substrate as a global back gate), shown in figure 2a. As it can be seen, there is a change in the magnitude of the hysteresis loop along with a changing slope of the linear background in the Hall resistance for changing $V_{BG}$. After removing a smooth polynomial background, the step size $\Delta R_{xy}$ (assigned as the amplitude of AHE) of the hysteresis loop can be obtained. The result is plotted as a function of the $V_{BG}$ in figure 2b showing a tunability in the AHE amplitude from ~4 Ω to ~11 Ω. The longitudinal resistance $R_{xx}$ (measured at the minimum of the magnetoresistance) is also plotted as a function of $V_{BG}$ in the same figure and a correlation with $\Delta R_{xy}$ is observed. In figure S4 the $\Delta R_{xy}$ is plotted against $R_{xx}$ and is fitted with a power law: $\Delta R_{xy} \sim R_{xx}^b$ with b~2.8±0.3. Following the literature [Zou2022, Jiang2015, Wang2015] on various microscopic mechanisms of the AHE and for our regime of $R_{xy}$<< $R_{xx}$, skew scattering would give an exponent b~1 whereas the side-step or Berry phase (intrinsic) as the mechanism for AHE would give b~2. Our observed power law (given the notable uncertainty in the data) appears to be closer to the expectation from the latter mechanisms and tends to rule out the skew-scattering as the dominant mechanism in our case. The extracted Hall coefficient (linear slope of $R_{xy}$ vs. magnetic field, reflecting the ordinary Hall effect) is plotted in figure 2c, and is clearly modulated (both the sign and amplitude) by the back gate. The change in the sign indicates the tunability of the dominant carrier type in the device from holes to electrons (note the discussions generally hold even in the case that our BSTS flake consist of two independent surface channels, with the bottom channel tuned more by the back gate). The amplitude of the Hall coefficient achieved in both carrier regimes would correspond to a carrier concentration on the order of few $10^{12}\ cm^{-2}$, similar to previous BSTS devices in the literature [Chong2019, Chong2018, Xu2014, Fatemi2014].

The fact that the hysteresis loop maintains its direction, or that the AHE ($\Delta R_{xy}$) does not change its sign, even when the Hall coefficient changes the sign or dominant carrier type is changed is a significant observation, as it rules out the trivial Hall effect due to any stray magnetic field produced by the ferromagnet (CGT) as a possible cause for the $R_{xy}$ hysteresis (in the trivial Hall effect scenario the hysteresis or $\Delta R_{xy}$ would also change sign when the main carrier type or Hall slope changes sign) [Jiang2015, Wang2015]. Furthermore, we have estimated the strength of the fringe magnetic field in such devices (no more than 1 mT [Kim2019]) is too weak to produce the observed $\Delta R_{xy}$ (of ~10 Ω, which would require a change in magnetic field on the order 0.1 T, as seen in Figure 1d).

We also studied the temperature dependence of the AHE. Figure 3a shows the evolution of $R_{xy}$ as the temperature is increased, where the hysteresis loop (AHE amplitude $\Delta R_{xy}$) is seen to vanish between 38 K and 60 K. This is consistent with the Curie temperature ($T_C$~50K) of CGT flakes of comparable thickness measured with magneto-optics Kerr effect in previous work [Idzuchi2019]. In figures S5 and S6 we further show the temperature dependence of the AHE in additional devices along with magneto optics Kerr effect (MOKE) hysteresis loops measured at the same time, showcasing that the AHE behavior measured by the magneto transport is consistent with the CGT magnetization probed by MOKE. The temperature dependence of the longitudinal resistance $R_{xx}$ is shown in figure S7. The observed metallic behavior is consistent our previous measurement of similar BSTS flakes (whose conduction is dominated by the metallic surface state at such temperatures) [Xu2014]. It is worth noting that our BSTS flake is only partially covered by CGT on the top surface, therefore the unmagnetized bottom surface and unmagnetized portion of the top surface are expected to make important contribution to $R_{xx}$.

In conclusion, we demonstrated that thin flakes of the topological insulator BSTS can show anomalous Hall effect when interfaced with thin flakes of CGT using simple mechanical dry transfer methods for the assembly. The observed AHE shows a rectangular hysteresis loop with coercive field of ~0.035 T at 2K and vanishes close to the $T_C$ of thin CGT flakes. Furthermore, the AHE can be tuned from ~4 to ~11 Ω and follows the longitudinal resistance of the device as a back gate is applied. These results indicate that simple mechanical transfer of van der Waals materials allows for magnetic proximity effect, which is the basis to realize novel magnetic topological phases such as QAHI. Our work may facilitate interfacing the growing inventory of layered materials with different magnetic or electrical properties.

**Experimental methods**.

*Device fabrication*. BSTS and CGT thin flakes were exfoliated on $SiO_2$/Si substrates inside a glovebox, where the oxygen and water concentration were less than 5 ppm. We chose appropriate CGT thin flakes (thickness between 4 nm – 10 nm) and BSTS flakes (thickness between 40 nm – 70 nm) based on the optical contrast with the thickness later confirmed by atomic force microscopy. We transferred the freshly exfoliated CGT flakes using the dry transfer technique on top of fresh BSTS flakes. Briefly, we used a PDMS/polycarbonate (PC) stamp to transfer the CGT flake from its initial substrate to the top of the chosen BSTS flake, the CGT flake is dropped along with the PC carrier film. To remove the PC film the sample is rinsed with chloroform for 10 minutes followed by acetone and isopropyl alcohol solvent cleaning. Then, the heterostructure is patterned using electron beam lithography. Finally, 5 nm of Cr and 95 nm of Au are deposited as contacts using e-beam evaporation. The prepared devices are then mounted for electrical transport measurements as quickly as possible after liftoff, usually with no more than half to an hour of exposure to air to avoid degradation of CGT.

*Electrical transport measurements.* Electrical transport measurements are performed in a variable temperature insert (VTI) system, which allows temperatures from 1.8K to 300K and an applied magnetic field of up to 6 T. The longitudinal resistance and Hall resistance were measured using four probe method with a Stanford Research SR830 lock-in amplifier with a low frequency (~13 Hz) excitation current of 100 nA – 1µA, or by applying 1-10 µA of DC current with a Keithley 2400 source meter and measuring the voltage drop with a Keithley 2182A nanovoltmeter. The gate control was achieved by applying a DC voltage to the p-doped silicon substrate using a Keithley 2400 source meter. The magnetic field dependent Hall resistance traces were antisymmetrized between traces measured with decreasing and increasing magnetic field (i.e., $R_{xy\ antisym}(B) = (R_{xy\ B\ decreasing}(B) - R_{xy\ B\ increasing}(-B))/2$). This is done to remove some field-even features from the Hall effect traces resulting from mixing of $R_{xx}$ into the $R_{xy}$ due to the geometry of the device and misalignment of electrodes. It is worth noting that the hysteresis loops and AHE modulation with back gate are still evident without the antisymmetrization step.

**Supplementary material**

See supplementary material for the thickness dependence of the hysteresis loops of the magnetization of CGT (as probed with MOKE, figure S1), temperature dependence of current in bare CGT indicating its insulating nature at low temperatures (figure S2), the HLN equation fits to the magnetoresistance (figure S3), power law fit of AHE amplitude vs. $R_{xx}$ (figure S4), AHE in other devices and their temperature dependence (figure S5), comparison of MOKE and AHE measured on the same CGT/TI device (figure S6) and the temperature dependence of $R_{xx}$ (figure S7).

**Acknowledgments**

We acknowledge P. Upadhyaya, J. Liao, G. Cheng, and J. Ribeiro for fruitful discussions. Different stages of this research at Purdue have been supported in part by National Science Foundation Emerging Frontiers & Multidisciplinary Activities (EFMA #1641101) and US Department of Energy (DOE) Office of Science through the Quantum Science Center (QSC, a National Quantum Information Science Research Center). Work at Tohoku has been supported in part by JSPS KAKENHI (Grants No. 17K14329, No. 18H04471, No. 18H04304, No.18F18328, and No. 18H03858) and the thermal management program of CREST, JST, research grants from The Iwatani Naoji Foundation's Research Grant, and AIMR with the support of World Premier International Research Center Initiative (WPI) from MEXT. Xing-Chen Pan acknowledges the support from an International Research Fellowship of Japan Society for the Promotion of Science [Postdoctoral Fellowships for Research in Japan (Standard)].

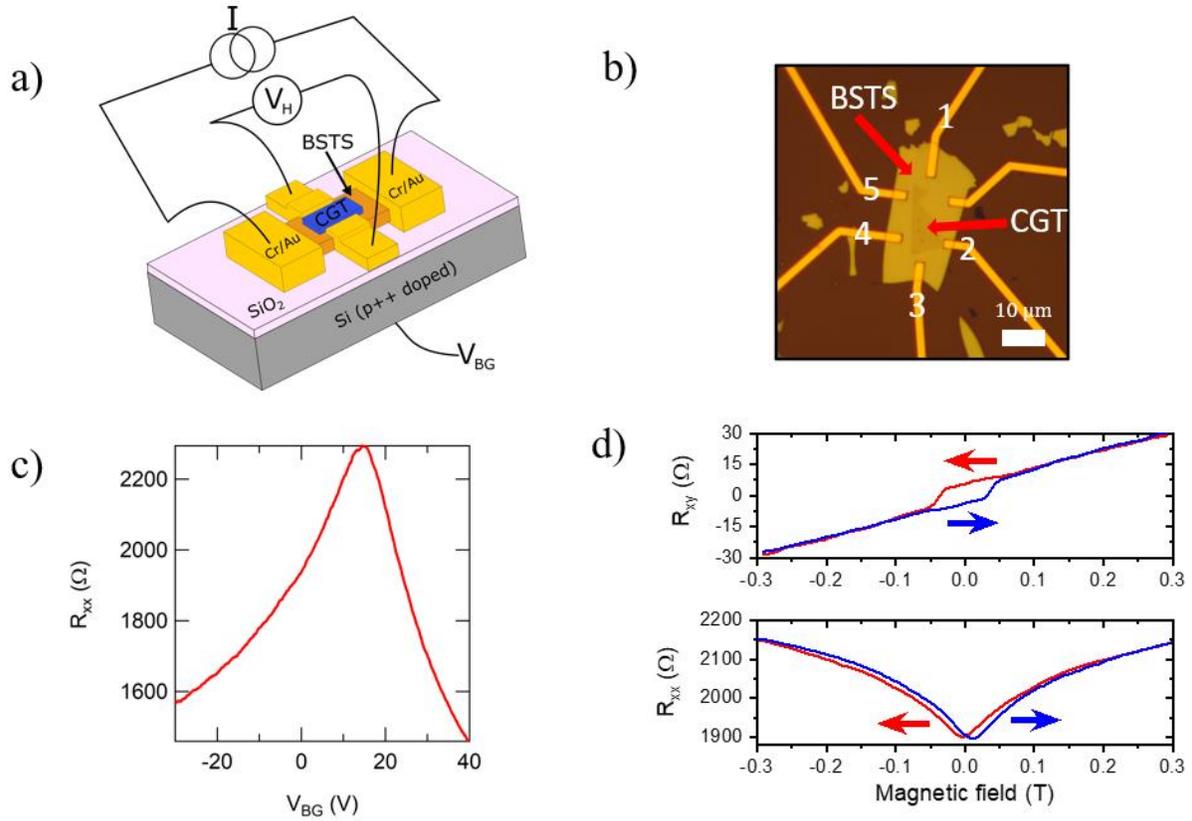

Figure 1. (a) Schematic of the $Cr_2Ge_2Te_6/BiSbTeSe_2$ (CGT/BSTS) back gated device (not showing all electrodes, only the ones injecting the current *I* and measuring the Hall voltage $V_H$ are shown). (b) Optical microscope image of a device, numbers indicate the electrodes used to measure $R_{xx}$ (5 to 4) and $R_{xy}$ (2 to 4) and inject current (1 to 3) for the 4-terminal measurements. CGT thickness is 6 nm and BSTS thickness is 66 nm. Scale bar is 10 µm. (c) Longitudinal resistance ($R_{xx}$) as a function of the back-gate voltage ($V_{BG}$) at T= 2 K and B=0 T. Ambipolar field effect is observed. (d) Hall resistance ($R_{xy} = V_H/I$) and longitudinal resistance ($R_{xx}$) measured at ~2 K and $V_{BG} = 0$ V as functions of an out-of-plane magnetic field (B), showing clear hysteresis. Correspondingly colored arrows indicate magnetic field sweep directions.

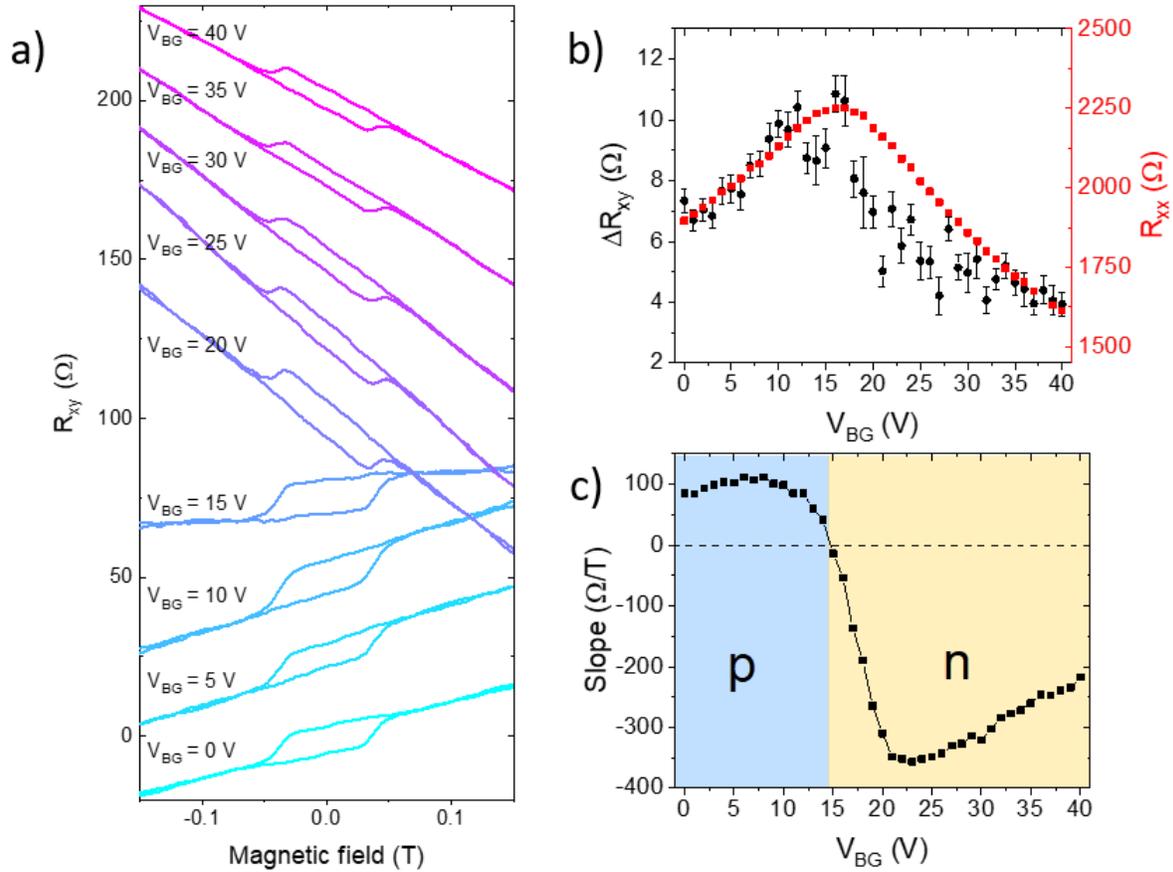

Figure 2. (a) Magnetic field dependent hall resistance ($R_{xy}$) measured for different back-gate voltages ($V_{BG}$) for the device shown in figure 1. Two traces labeled by their $V_{BG}$ correspond to opposite magnetic field sweeps (the trace with higher/lower $R_{xy}$ at the hysteresis loop is measured with decreasing/increasing magnetic field). All measurements done at T = 2 K with an excitation DC current of 10 µA. Traces are offset for clarity. (b) Amplitude of the resistance hysteresis due to anomalous Hall effect, $\Delta R_{xy}$ (obtained after subtracting a polynomial background in magnetic field dependent $R_{xy}$ data, see Supplemental Fig. S3 for an example), plotted on the left axis as a function of $V_{BG}$. The $R_{xx}$ data (right axis) were extracted from the average of the $R_{xx}$ minima in the two traces with opposite magnetic field sweeps (such as those in Fig. 1d lower panel). (c) Slope (related to the ordinary Hall effect) of the linear part of the background in the $R_{xy}$ vs. magnetic field, plotted as a function of $V_{BG}$. The sign change indicates a change in the dominant type of carriers.

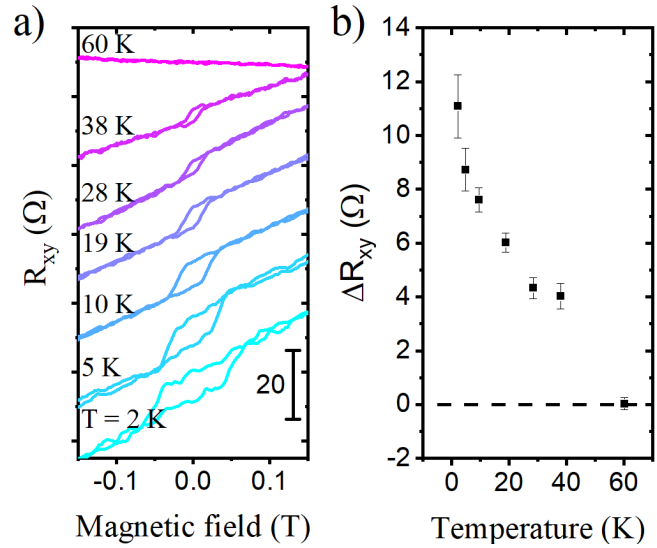

Figure 3. (a) Hall resistance ($R_{xy}$ vs magnetic field) for different temperatures (traces offset for clarity) and (b) hysteresis amplitude related to the anomalous Hall effect (right, $\Delta R_{xy}$) as a function of temperature. Dashed line marks the zero level.

**Supplementary information:**

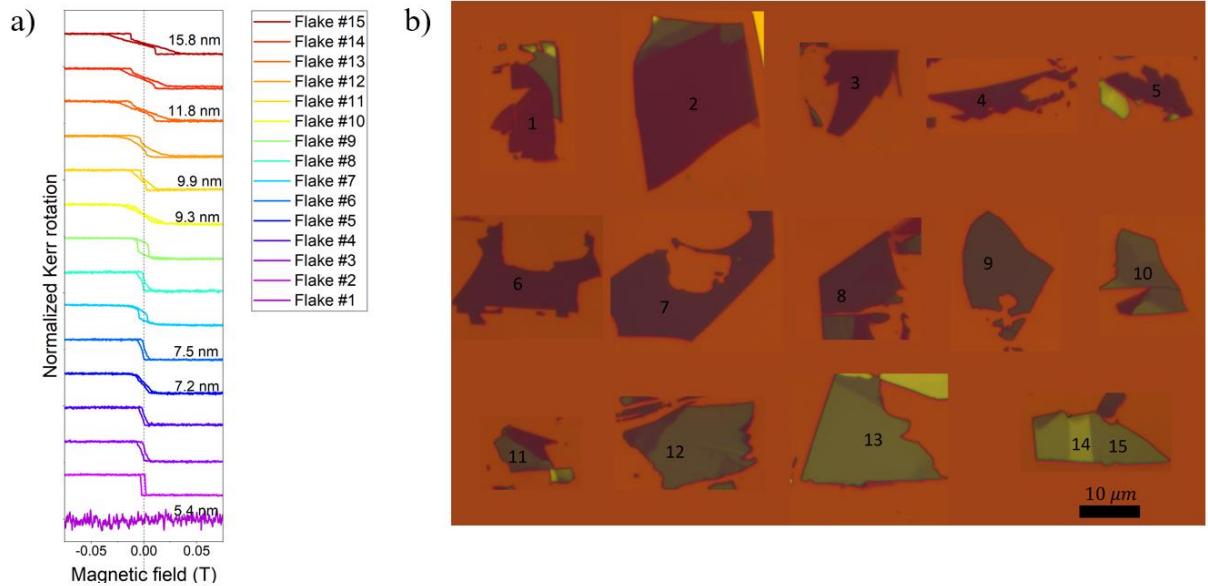

Figure S1. A) Magneto optical Kerr effect (MOKE) characterizations for bare CGT flakes of different thicknesses. b) Optical micrographs of the measured flakes ordered according to approximate thickness (as estimated from color in optical micrographs) increasing from top-left to bottom-right. Measurement temperature: 10 K. Below 9-10 nm (but above ~5nm) the MOKE curves exhibit more rectangular hysteresis loops and above 10 nm hysteresis loops have more structures usually attributed to vortex or labyrinth type of domains formation. For thin enough flakes (<5 nm) no MOKE signal is seen attributed to sample degradation.

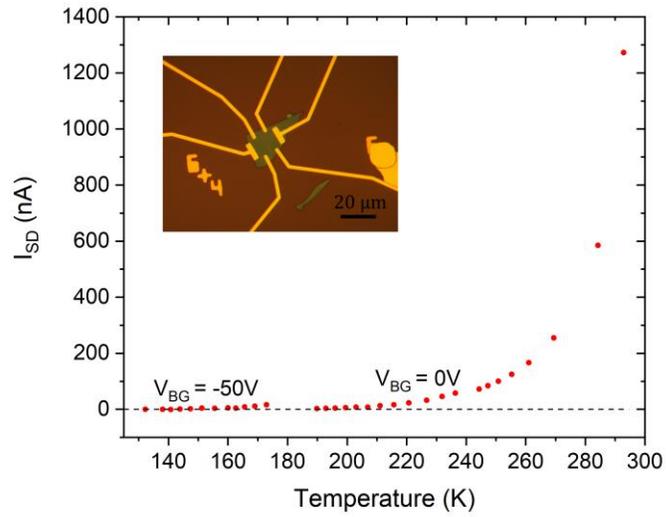

Figure S2. Source-drain current as a function of temperature for a bare CGT flake. Source-drain voltage is 10 V and measurement is done in the two-terminal configuration, below 200K the resistance is above 1 GΩ. For the 130 K-170 K temperature range the back gate is -50 V. In general, applying negative back gate voltage increased the source-drain current and positive gate voltage reduced the source-drain current indicating a p-doped CGT. Inset is an optical micrograph of the device; scale bar is 20 μm. We use the same type of contacts as BSTS/CGT devices: Cr/Au (5 nm/95 nm).

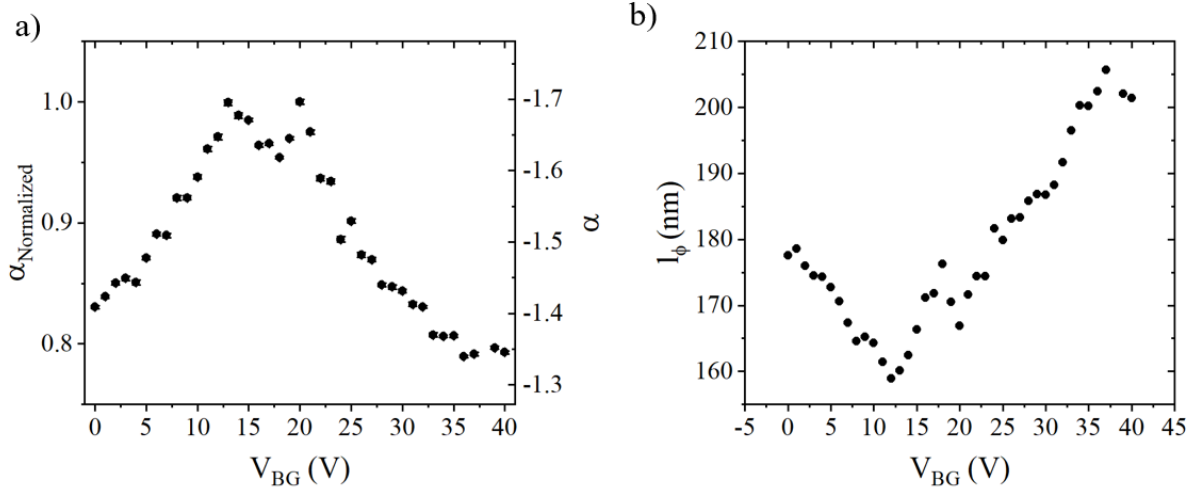

Figure S3. The Hikami-Larkin-Nagaoka (HLN) equation describing the WAL, $\Delta\sigma(B) = -\alpha \frac{e^2}{2\pi^2\hbar}\left[\ln\left(\frac{B_0}{B}\right) - \Psi\left(\frac{1}{2} + \frac{B_0}{B}\right)\right]$, with $B_0 = \frac{\hbar}{4el_\phi^2}$, was fitted to the longitudinal magnetoresistance ($R_{xx}(B)$) for different gate voltages. a) The fitting parameter $\alpha$ as a function of back gate voltage. Due to the geometry of the device, there is some uncertainty in extracting the amplitude of the sheet conductance and the absolute amplitude of $\alpha$ (subject to an unknown prefactor of order 1). Therefore we also plot the normalized (by the peak value) $\alpha$ on the left and mainly focus on the trend of gate dependence of $\alpha$, showing the absolute value of $\alpha$ generally increases around the Dirac point ($V_{BG} \sim 15$ V) compared to far away from Dirac point, consistent with the interpretation of $\alpha$ as the number of independent coherent channels (from surfaces and bulk), which tend to be more decoupled when gated close to Dirac point (thus increasing $|\alpha|$) [Steinberg2011, Jauregui2015]. b) The back gate dependence of the other fitting parameter $l_\phi$.

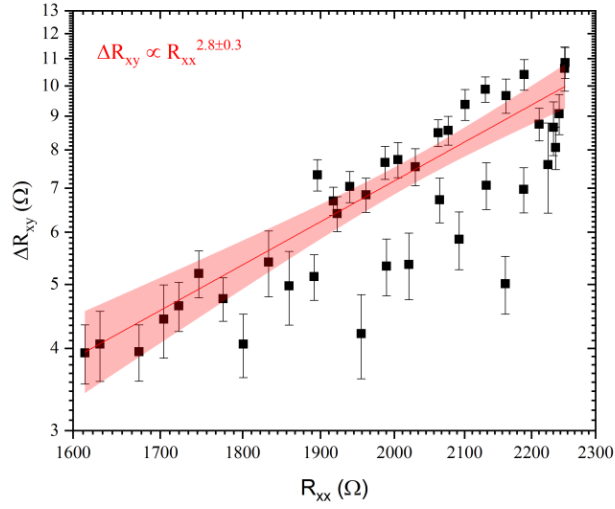

Figure S4. Double logarithmic plot of $\Delta R_{xy}$ as a function of $R_{xx}$ (as extracted from figure 2b), A power law with $\Delta R_{xy} \sim R_{xx}^b$ is extracted, $b = 2.8 \pm 0.3$.

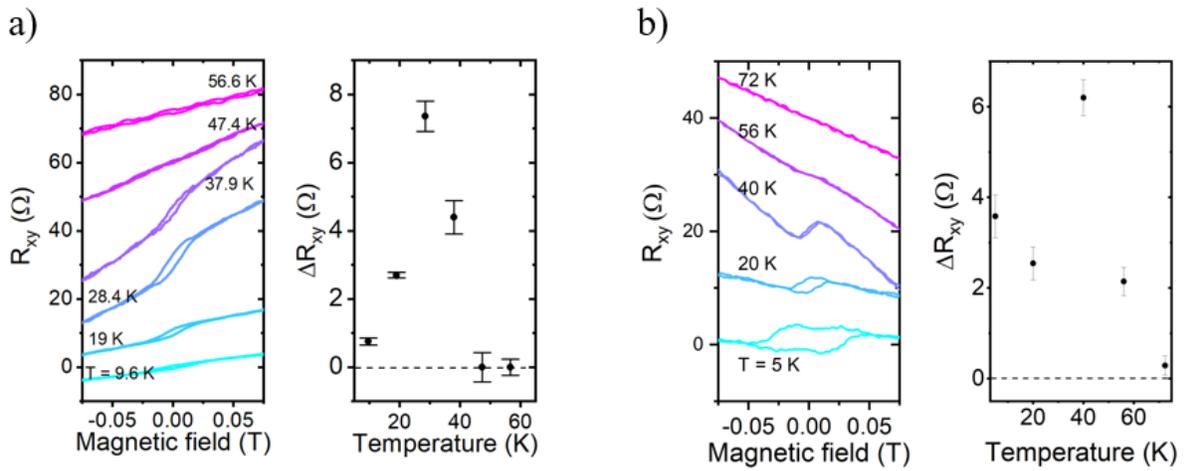

Figure S5. $R_{xy}$ traces vs magnetic field ($R_{xy}$, left) and temperature dependence of the hysteresis amplitude related to the AHE ($\Delta R_{xy}$, right) for two other CGT/BSTS devices. (a) BSTS thickness ~43 nm, CGT thickness ~4 nm. (b) BSTS thickness ~55 nm, CGT thickness ~8 nm.

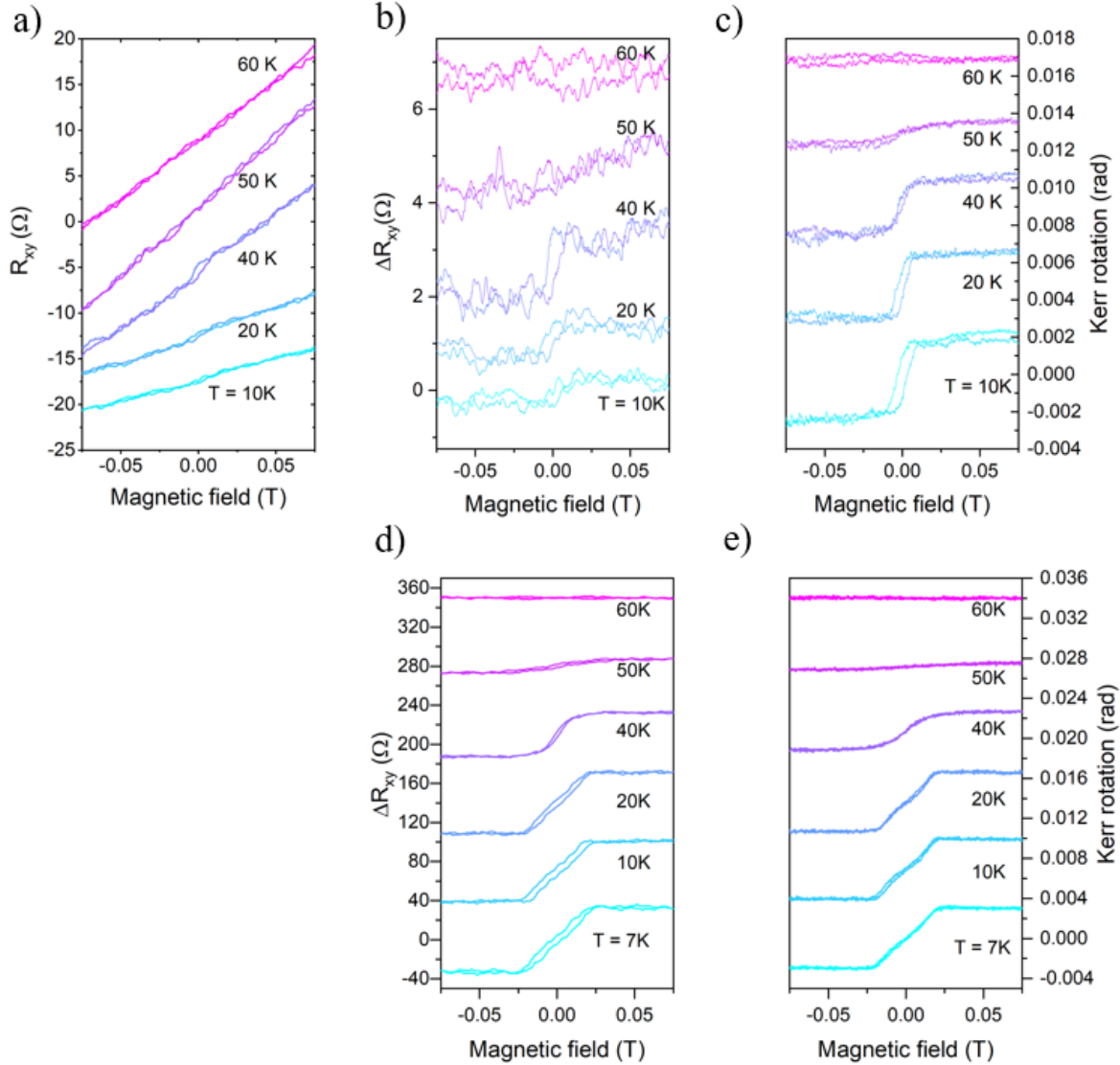

Figure S6. (a) $R_{xy}$ and (b) $\Delta R_{xy}$ ($R_{xy}$ with linear background removed) vs magnetic field at different temperatures for a device, with concurrently measured MOKE from the same device (c). For MOKE, a 635 nm laser with excitation power less than 3 µW was used. Transport and MOKE data were taken at the same time. BSTS thickness ~55 nm, CGT thickness ~8 nm. Under laser excitation the AHE was reduced (see figure S5b, the same device measured in the VTI system where only transport measurements were performed). This could be due to a photodoping effect and/or aging of the device. A change in carrier type is also observed when compared to the initial measurement in figure S5b. (d) AHE (left) and (e) MOKE (right) measured at the same time for another device with CGT (thickness ~15 nm) on the bottom and relatively thin BSTS (thickness ~15 nm) on top. This device in particular showed stronger AHE ($\Delta R_{xy}$ ~80 $\Omega$). Traces in each panel vertically offset for clarity.

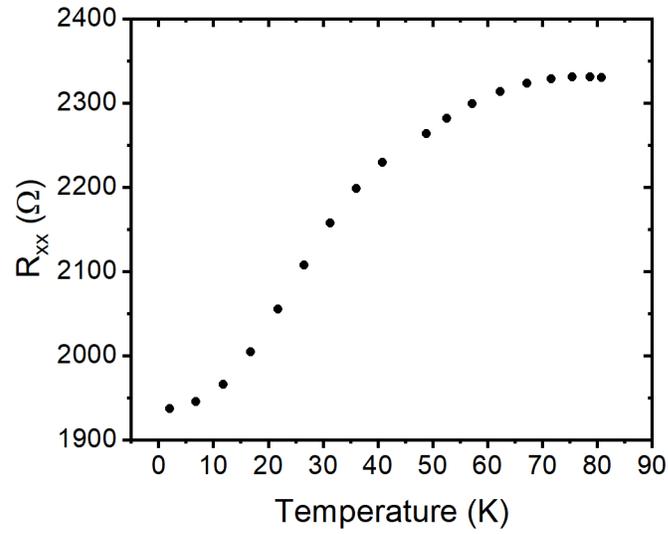

Figure S7. Temperature dependence of the longitudinal resistance at zero magnetic field for the device shown in figure 1.